\documentstyle[aps,eqsecnum,epsf]{revtex}
\newcommand{\be}{\begin{equation}}
\newcommand{\ee}{\end{equation}}
\newcommand{\bea}{\begin{eqnarray}}
\newcommand{\eea}{\end{eqnarray}}
\newcommand{\pas}{/ \kern-0.55em\partial}
\newcommand{\ps}{/ \kern-0.55em p}
\newcommand{\Ds}{/ \kern-0.69em D}
\newcommand{\ks}{/ \kern-0.55em k}
\begin{document}
\voffset = 0.5in 
\draft
\title{Quantum Electrodynamics with the Pauli Term}
\author{Ramchander R. Sastry}
\address{Center for Particle Physics,\\
 University of Texas at Austin, \\
   Austin, Texas 78712-1081.}
\date{\today}
\maketitle

\begin{abstract}
The quantum field theory of extended objects is employed to address the
hitherto nonrenormalizable Pauli interaction.  This is achieved by quantizing
the Dirac field using the infinite dimensional generalization of the 
extended object formulation following which the order $\alpha$ contribution to the anomalous magnetic moment of the electron (and of the muon) arising from the Pauli term is calculated.
\end{abstract}


\section{Introduction}
The quantum mechanics of extended objects\cite{sastry1} and its infinite dimensional generalization, namely, the quantum field theory of extended objects, in
particular $\phi^6$ scalar field theory have been presented by the author
\cite{sastry2}.  The quantum field theory of extended objects is obtained by
constructing the Lorentz invariant generalization to fields of the noncanonical
commutation relations
\be
\label{co-eqn}
\left[X_{f_{\mu}},P_{\nu}\right] = ie^{-P^2/m^2}g_{\mu\nu}
\ee
where $X_{f_{\mu}}$ is the noncommuting fuzzy 4-position operator, $P_{\nu}$ is the Euclidean 4-momentum, and $g_{\mu\nu} = (1,1,1,1)$ is the Euclidean metric.  The commutation relations in Eq.~(\ref{co-eqn}) have a smooth limit with ordinary quantum mechanics as the Compton wavelength $\frac{1}{m}$ of the particle vanishes.   We choose to construct the field theory in Euclidean space for convenience.  As we shall demonstrate the Minkowski space propagation amplitude can be recovered by the variable transformation $p_0 \rightarrow ip_0$.  This is not a rotation of the contour in the sense of the Wick rotation since the contour does not close in this case.  Such a transformation ensures that the preservation of the four dimensional rotational symmetry of Euclidean space will also preserve Lorentz invariance.  After quantizing the Dirac field we study quantum electrodynamics with the Pauli term.  The order $\alpha$ contribution to the magnetic moment of the electron (and of the muon) arising from the hitherto nonrenormalizable correction term is calculated.   

\section{The Dirac Field}
The Euclidean space Lagrangian for the free Dirac field can be written as
\be
{\cal L} = {\overline \psi}
(i\pas + m)\psi
 = {\overline \psi}(i\gamma^{\mu}\partial_{\mu} + m)\psi.
\ee   
The Dirac field is usually quantized by imposing the anticommutation relations
\be
\left\{\psi_a^{(3)}({\bf x}),\psi_b^{(3)\dagger}({\bf y})\right\}
 = \delta^{(3)}({\bf x} - {\bf y})\delta_{ab}
\ee
where $\psi^{(3)}(x)$ is expanded as a sum over 3-momenta:
\be
\phi^{(3)}(x) = \int \frac{d^{3}p}{(2\pi)^{3}}\frac{1}{\sqrt{2E_{{\bf p}}}}
\sum_s(a_p^su^s(p)e^{-ipx} + b_p^sv^s(p)e^{ipx})
\ee
and similarly for $\psi^{(3)}(x) = \psi^{(3)\dagger}(x)\gamma^0$.  Here $a$ and $b$ are the annihilation operators for particles and antiparticles obeying the anticommutation relations
\be
\left\{a_{{\bf p}}^r,a_{{\bf q}}^{s\dagger}\right\} = \left\{a_{{\bf p}}^r,
a_{{\bf q}}^{s\dagger}\right\} = (2\pi)^3\delta^{(3)}({\bf p} - {\bf q}).
\ee
The spinor contraction relations in Euclidean space are given by
\bea
\sum_su^s(p){\overline u}^s(p) &=& \gamma \cdot p - m\\ \nonumber
\mbox{and}\\
\sum_sv^s(p){\overline v}^s(p) &=& \gamma \cdot p + m. \nonumber
\eea
The retarded Dirac field propagation amplitude can be computed from the above
expansions and is given by
\bea
\label{3-eqn}
S_R^{ab}(x - y) &=& \theta(x^0 - y^0)\langle 0|\left\{\psi_a^{(3)}(x),
{\overline \psi^{(3)}}_b(y)\right\}|0\rangle\\ \nonumber
&=& \int \frac{d^{4}p}{(2\pi)^{4}}\frac{(\ps - m)}{p^{2} + m^{2}}
e^{-ip(x - y)} 
\eea
where the anticommutator is employed in order to ensure that the vacuum has only positive-energy excitations.  The same propagation amplitude can be obtained by expanding the field as a
sum over 4-momenta:
\be
\psi(x) = \int\frac{d^{4}p}{(2\pi)^{4}}\frac{1}{\sqrt{p^{2} + m^{2}}}
\sum_s(a_p^su^s(p)e^{-ipx} + b^s_pv^s(p)e^{ipx}).
\ee
The field expansion for $\psi(x)$ coupled with the Euclidean spinor field anticommutation relations:
\be
\left\{\psi_a(p),\psi_b^{\dagger}(q)\right\} = i\delta^4(p - q)\delta_{ab}
\ee
which are Lorentz invariant generalizations of the commutation relations arising from ordinary (spacetime) quantum mechanics (zero Compton wavelength limit of Eq.(\ref{co-eqn})) leads to 
\bea
\label{prop-eqn}
S_R^{ab}(x - y) &=& \langle 0|\psi(x){\overline \psi}(y)|0\rangle\\ \nonumber
                &=& \int \frac{d^{4}p}{(2\pi)^{4}}\frac{(\ps - m)}
                    {p^{2} + m^{2}}e^{-ip(x - y)}
\eea 
which is the Dirac field propagation amplitude.  We observe that by changing
$p \rightarrow -p$ the right hand side of Eq.(\ref{prop-eqn}) is symmetric
under $x \leftrightarrow y$ implying that $\left[\psi_a(x),{\overline \psi}_b
(y)\right] = 0$.  The vanishing of the commutator can be explained by noting
that $\psi(x)$ is a field which creates and destroys states characterized by
relativistically invariant 4-momenta (or mass when on shell).  Thus, the
measurement of a field which creates and destroys such states at one spacetime point cannot affect its measurement at another spacetime point.  We note that $\psi(x)$ is not the usual 3-momentum expansion of the Dirac field (denoted by $\phi^{(3)}(x)$).  From Eq.~(\ref{3-eqn}) and Eq.~(\ref{prop-eqn}) it follows that 
\be
\langle 0|\psi(x){\overline \psi}(y)|0\rangle = \theta(x^0 - y^0)\langle 0|
\left\{\psi_a^{(3)}(x),{\overline \psi^{(3)}}_b(y)\right\}|0\rangle.  
\ee
The anticommutator on the right hand side vanishes for spacelike separations and it is well known that the ordinary Dirac field preserves causality.  This is not surprising since (as shown above) the propagator can also be obtained by a 
Lorentz invariant generalization of a causal structure, namely, ordinary (spacetime) quantum mechanics.  The reason this approach is not employed in the quantum field theory of point particles is because the existence of point particle mass states violates the statement of causality\cite{sastry2}.

In order to characterize the intermediate states in a nonrenormalizable theory we generalize the noncanonical commutation relations arising from the extended object formulation:
\be
\left\{\psi_a(p),\psi_b^{\dagger}(q)\right\} = ie^{-p^2/m^2}\delta^4(p - q)
\delta_{ab}
\ee
where, once again, the anticommutator is employed in order to ensure that the
vacuum has only positive-energy excitations.  As a consequence, the propagation amplitude becomes
\bea
\langle 0|\psi(x){\overline \psi}(y)|0\rangle
             = \int \frac{d^{4}p}{(2\pi)^{4}}\frac{(\ps - m)}
               {p^{2} + m^{2}}e^{-p^2/m^2}e^{-ip(x - y)}
\eea 
which is once again symmetric under $x \leftrightarrow y$ and reflects the fact that relativistically invariant 4-momentum states are being created and
destroyed by the field $\psi(x)$.  We note that the imposition of noncanonical commutation relations precludes the existence of 3-momentum characterizations for the particle.  The author has proved this statement for the scalar field and the extension to the Dirac field is straightforward\cite{sastry2}.  This fact coupled with the fact that we are employing a Lorentz invariant generalization of a causal structure, namely, the extended object formulation, allows us to conclude that the statement of microscopic causality is not violated by this quantization procedure.  We observe the the Euclidean momentum space propagator
\be
{\tilde S}(p) = \frac{(\ps - m)}{p^{2} + m^{2}}e^{-p^2/m^2}
\ee
is bounded from above and below since $p^2$ is a Euclidean scalar.  The ordinary Dirac field propagator (Euclidean) is recovered in the limit of vanishing Compton wavelength.  A crucial feature of this propagator is the Gaussian damping which eliminates the high frequency modes and renders the interacting theory finite to all orders.  The Minkowski space propagation amplitude can be obtained by the variable transformation $p_0 \rightarrow ip_0$ which is not a rotation of the contour in the sense of the Wick rotation since the contour does not close due to the essential singularity at infinity.  The corresponding $dp_0$ integration limits in the propagation amplitude go from $-i\infty$ to $i\infty$.  In order to explain this feature we observe that the spacetime commutation relations arising from the extended object formulation have a relative negative sign between the phase space and energy-time relations.  This sign difficulty can be removed by using the Minkowski metric but this does not afford a Lorentz invariant generalization to fields.  Thus, if we choose to formulate the theory in Minkowski space we have to choose our $dp_0$ integration limits from $-i\infty$ to $i\infty$.  We note that if the contour were to close this would simply be a Wick rotation as is the case in the limit of vanishing Compton wavelength.  In his previous paper on $\phi^6$ scalar field theory the author has proved that unitarity is preserved in the theory up to fourth order in the coupling constant\cite{sastry2}.  In this paper we do not prove unitarity in the theory but appeal to the argument that since the scalar field propagator preserves unitarity in a nonrenormalizable theory, namely, $\phi^6$ theory, its spinor extension will also preserve unitarity in a corresponding nonrenormalizable theory, namely, quantum electrodynamics with the Pauli term.  Moreover, we can subject the calculated values of the order $\alpha$ corrections arising from the Pauli term to experimental verification.

\section{The Pauli Interaction}
Following the quantization of the Dirac field we proceed to calculate the order $\alpha$ correction to the magnetic moment contribution arising from the Pauli term.  In Minkowski space with $\eta_{\mu\nu} = (-1,1,1,1)$ the Lagrangian for quantum electrodynamics with the Pauli interaction is:
\be    
{\cal L} = {\overline \psi}(i\Ds  - m)\psi - \frac{1}{4}F_{\mu\nu}
F^{\mu\nu} - \frac{e}{M}{\overline \psi}\sigma_{\mu\nu}\psi F^{\mu\nu}
\ee
where $D_{\mu}$ is the gauge covariant derivative
\be
D_{\mu} = \partial_{\mu} + ieA_{\mu}(x),
\ee
$F_{\mu\nu} = \partial_{\mu}A_{\nu} - \partial_{\nu}A_{\mu}$ is the electromagnetic field strength tensor, $\sigma^{\mu\nu} = \frac{i}{2}\left[\gamma^{\mu},
\gamma^{\nu}\right]$, and $M$ is some common mass.  The last term in the Lagrangian is the Pauli term which is the leading nonrenormalizable correction term of dimension $5$ allowed by CP, Lorentz, and gauge invariance.  We will now study the order $\alpha$ correction to electron scattering due to the presence of a virtual photon.  The vertex correction diagram is shown in figure 1 and the S-matrix element for scattering from this field is given by
\be
i{\cal M}(2\pi)\delta(p'_0 - p_0) = -ie{\overline u}(p')\left[\Gamma^{\mu}
{\tilde A}^{cl}_{\mu}(p' - p) + \frac{1}{M}\Gamma^{\mu\nu}{\tilde F}_{\mu\nu}^{cl}(p' - p)\right]u(p)
\ee
which contains the sum of two contributions.  The first contribution is the
conventional vertex correction arising due to the interaction
\be
\Delta H_{int} = \int d^3x eA_{\mu}^{cl}j^{\mu}
\ee
where $j_{\mu}(x) = {\overline \psi}(x)\gamma^{\mu}\psi(x)$ is the electromagnetic current and $A_{\mu}^{cl}$ is a fixed classical potential.  The second contribution is the vertex correction arising due to the Pauli interaction
\be
\Delta H_{int} = \int d^3x \frac{e}{M}F_{\mu\nu}^{cl}j^{\mu\nu}
\ee 
where $j_{\mu\nu}(x) = {\overline \psi}(x)\sigma^{\mu\nu}\psi(x)$ is the Pauli tensor current and $F_{\mu\nu}^{cl}$ is a fixed classical field.  The first contribution leads to the well known anomalous magnetic moment first calculated by Schwinger in 1948\cite{schwinger}.  The Lande $g$-factor is corrected (at order $\alpha$) by an amount
\be
a_e = \frac{g - 2}{2} = \frac{\alpha}{2\pi} \approx 0.0011614
\ee
whereas the experimental value is $a_e = 0.001159652209(31)$ where the estimated errors are in parentheses\cite{combley}.  Since the time the calculation was originally performed by Schwinger, the calculation has been carried to order $\alpha^3$ (where there are $72$ Feynman diagrams).  The theoretical value to this order is given by:\cite{kaku}
\be
a_e = 0.5\left(\frac{\alpha}{\pi}\right) - 0.32848\left(\frac{\alpha}{\pi}\right)^2 + 1.49\left(\frac{\alpha}{\pi}\right)^3 + \ldots
\ee
which leads to $a_e = 0.001159652411(66)$ where once again the estimated errors are in parentheses.  If we apply the extended object formulation to this problem we find that the resulting value for $a_e$ does not compare well with experiment.  This is because ordinary QED processes occur at low mass scales or equivalently at distance scales which are large.  Since we had defined the finite extent of a particle via its Compton wavelength which measures the distance over which quantum effects can persist, there is simply no coupling to the system for large distances.  Hence, the point particle formulation correctly gives the most accurate value for $a_e$. 

In order to study the order $\alpha$ correction due to the Pauli term (where the point particle approximation has to be relaxed due to the large mass scales) we need to evaluate the Feynman diagram shown in Figure 1.  We can restrict the form of $\Gamma^{\mu\nu}$ considerably by appealing to Lorentz invariance.  Since $\Gamma^{\mu\nu}$ transforms as a tensor it must be proportional to $\sigma^{\mu\nu}$ which is the only tensor available.  Therefore, we have
\be
\Gamma^{\mu\nu} = \sigma^{\mu\nu}\left[1 + G(q^2)\right]
\ee
where $G(q^2)$ is a form factor.  To lowest order $G = 0$ and $\Gamma^{\mu\nu} = \sigma^{\mu\nu}$.  By evaluating the vertex correction due to the Pauli term using the relations
\bea
{\overline u}(0)\left[\gamma^i,\gamma^j\right]u(0) &=& 
2i\epsilon^{}\sigma^k\\ \nonumber
\mbox{and}\\
{\overline u}(0)\left[\gamma^i,\gamma^0\right]u(0) &=& 0 \nonumber
\eea
and retaining terms up to first order in small momenta we observe that the Pauli term contributes an amount of order $\frac{4e}{M}$ to the magnetic moment of the electron.  The calculated value of the magnetic moment of the electron agrees with experiment to within terms of order $10^{-10}e/2m_e$ so $M$ must be greater than about $8\times 10^{10}m_e = 4\times10^{7}$ GeV\cite{weinberg}.  Applying the Feynman rules to the diagram in figure 1 we find, to order $\alpha$ that
\bea
\label{corr-eqn}
\lefteqn{\delta\Gamma^{\mu\nu}(p',p) = \int\frac{d^4k}{(2\pi)^4}
\frac{-ig_{\rho\delta}}
{(k - p)^2}{\overline u}(p')
\left(-i\frac{e}{M}\sigma^{\delta\alpha}\right)
\frac{i(\ks + m)}{k^2 + m^2}e^{-k^2/m^2} }\\
& &\hspace{1.5in}\times\:\left(-i\frac{e}{M}\sigma^{\mu\nu}\right) 
\frac{i(\ks ' + m)}{k'^2 + m^2}e^{-k'^2/m^2}  
\sigma^{\rho\alpha}u(p)
\eea
where $\Gamma^{\mu\nu} = \sigma^{\mu\nu} + \delta\Gamma^{\mu\nu}$ and 
$k' = k + q$.  We note that since we are dealing with retarded propagators the Feynman prescription for treating the poles is not employed.  By using the contraction identities
\bea
\gamma^{\nu}\gamma^{\mu}\gamma_{\nu} &=& 2\gamma^{\mu}\\ \nonumber
\gamma^{\rho}\gamma^{\mu}\gamma^{\nu}\gamma_{\rho} &=& -4g^{\mu\nu}\\ \nonumber
\gamma^{\rho}\gamma^{\mu}\gamma^{\nu}\gamma^{\sigma}\gamma_{\rho} 
&=& \gamma^{\sigma}\gamma^{\nu}\gamma^{\mu}
\eea
we can reduce Eq.~(\ref{corr-eqn}) to
\be
\delta\Gamma^{\mu\nu}(p',p) = \frac{-2ie^2}{M^2}
\int\frac{d^4k}{(2\pi)^4}
{\overline u}(p')(2kk' + m^2)\sigma^{\mu\nu}u(p)
\frac{1}{(k - p)^2)}\frac{e^{-k^2/m^2}}{k^2 + m^2}
\frac{e^{-k'^2/m^2}}{k'^2 + m^2}.
\ee
Employing the method of Feynman parameters we can express the denominator as
\be
\frac{1}{(k - p)^2}\frac{1}{k^2 - m^2}\frac{1}{k'^2 - m^2} = 
\int_0^1dx\,dy\,dz\,\delta(x + y + z - 1)\frac{2}{D^3}
\ee
where the new denominator $D$ is 
\bea
D &=& x(k^2 -m^2) + y(k'^2 - m^2) + z(k - p)^2\\ \nonumber
&=& k^2 + 2k\cdot(yq - zp) + yq^2 + zp^2 +(x + y)m^2.
\eea
By shifting $k$ to $l = k + yq - zp$ we obtain
\be
D = l^2 + \Delta
\ee
where
\be
\Delta = xyq^2 + (1 - z)^2m^2 > 0.
\ee
Consequently, we obtain a simplified expression for $\delta\Gamma^{\mu\nu}$:
\bea
\lefteqn{\delta\Gamma^{\mu\nu}(p',p) = \frac{-4ie^2}{M^2}
\int_0^1dx\, dy\, dz \, \delta(x + y + z - 1) } \\ 
& & \hspace{1in} \times \: \int\frac{d^4l}{(2\pi)^4} 
{\overline u}(p')(2kk' + m^2)\sigma^{\mu\nu}u(p)
\frac{e^{-(l - yq + zp)^2/m^2 -(l - yq + zp - q)^2/m^2}}
{(l^2 + \Delta)^3}.
\eea
As noted before the $dl_0$ integration limits go from $-i\infty$ to $i\infty$ since we have employed the noncanonical quantization procedure to obtain the propagator.  We can switch to Euclidean momenta by means of the variable transformation $l_0 \rightarrow il_0$ upon which we obtain the form factor $G(q^2)$ as
\bea
\lefteqn{G(q^2) = \frac{2\alpha}{\pi M^2}
\int_0^1dx\,dy\,dz\,\delta(x + y + z - 1) }\\
& & \hspace{1.0in}\times \:\int_0^{\infty} l^3dl \left[
2(l - yq + zp)(l - yq + zp - q) + m^2\right]
\\
& & \hspace{1.0in}\times \:
\frac{e^{-(l - yq + zp)^2/m^2 -(l - yq + zp - q)^2/m^2}}
{(l^2 + \Delta)^3}.
\eea
where the momenta are now Euclidean.  For a slowly varying electromagnetic field we can take the limit $q \rightarrow 0$ in the spinor matrix element and we obtain
\be
G(q^2 = 0) = \frac{2\alpha}{\pi M^2}
\int_0^1 dz\,\int_0^{1 - z}dy
\int_0^{\infty} l^5dl
\frac{2(l + zp)^2 + m^2}{(l^2 + \Delta)^3}e^{-2(l + zp)^2/m^2}.
\ee
 where we have increased the power of the momentum in the integrand by a factor of two in order to maintain the dimensionality of the graph.  This is due to the negative dimensional (mass dimension = -1) coupling constant.  In order to get an estimate of the value of $G(0)$ we can numerically compute the value of the integral at zeroth order in $p$.  If we choose the greatest lower bound for $M$ at $4\times10^7$ GeV, then the bound obtained for $G(0)$ is
\be
G(0) < 2.510402 \times 10^{-26}
\ee
This bound represents the order $\alpha$ correction to the Lande $g$-factor of the electron arising from the Pauli term.  The value of the order $\alpha$ correction is extremely small because of the smallness of the electron mass and the large bound on $M$.  The limit on $M$ can be weakened if other symmetries restrict the form of the nonrenormalizable interactions.  We observe that the full Lagrangian of quantum electrodynamics is invariant under the symmetry $\psi \rightarrow \gamma_5\psi$ (chiral transformation) and $m \rightarrow -m$.  If we impose this symmetry, the Pauli term would have to appear in the Lagrangian with an extra factor $m/M$, so that its contribution to the magnetic moment would be only of order $4em/M^2$.  Because of the extra factor of $m$, here it is the muon rather than the electron that provides the most useful limit on $M$.  The calculated value agrees with experiment to within terms of order $10^{-8}e/2m_e$, so $M$ must be greater than about $\sqrt{8\times 10^8
}m_{\mu} = 3 \times 10^3$ GeV\cite{weinberg}.  Using this value of $M$ and $m_{\mu} = 106$ MeV we obtain the bound on the order $\alpha$ correction (at zeroth order in $p$) to the Lande $g$-factor arising from the Pauli term (of the muon) as
\be
G_{\mu}(0) < 7.53197 \times 10^{-22}. 
\ee

\section{Conclusion}
By quantizing the Dirac field using the quantum field theory of extended objects we are able to calculate the order $\alpha$ correction to the anomalous magnetic moment contribution arising from the Pauli term.  The calculated values of the order $\alpha$ corrections arising from the Pauli term need to be subjected to experimental verification.

\acknowledgements
I would like to thank Rafal Zgadzaj for performing the numerical computations in Mathematica.

\newpage
\begin{figure}
\centerline{
\epsfxsize = 4.0in
\epsfbox{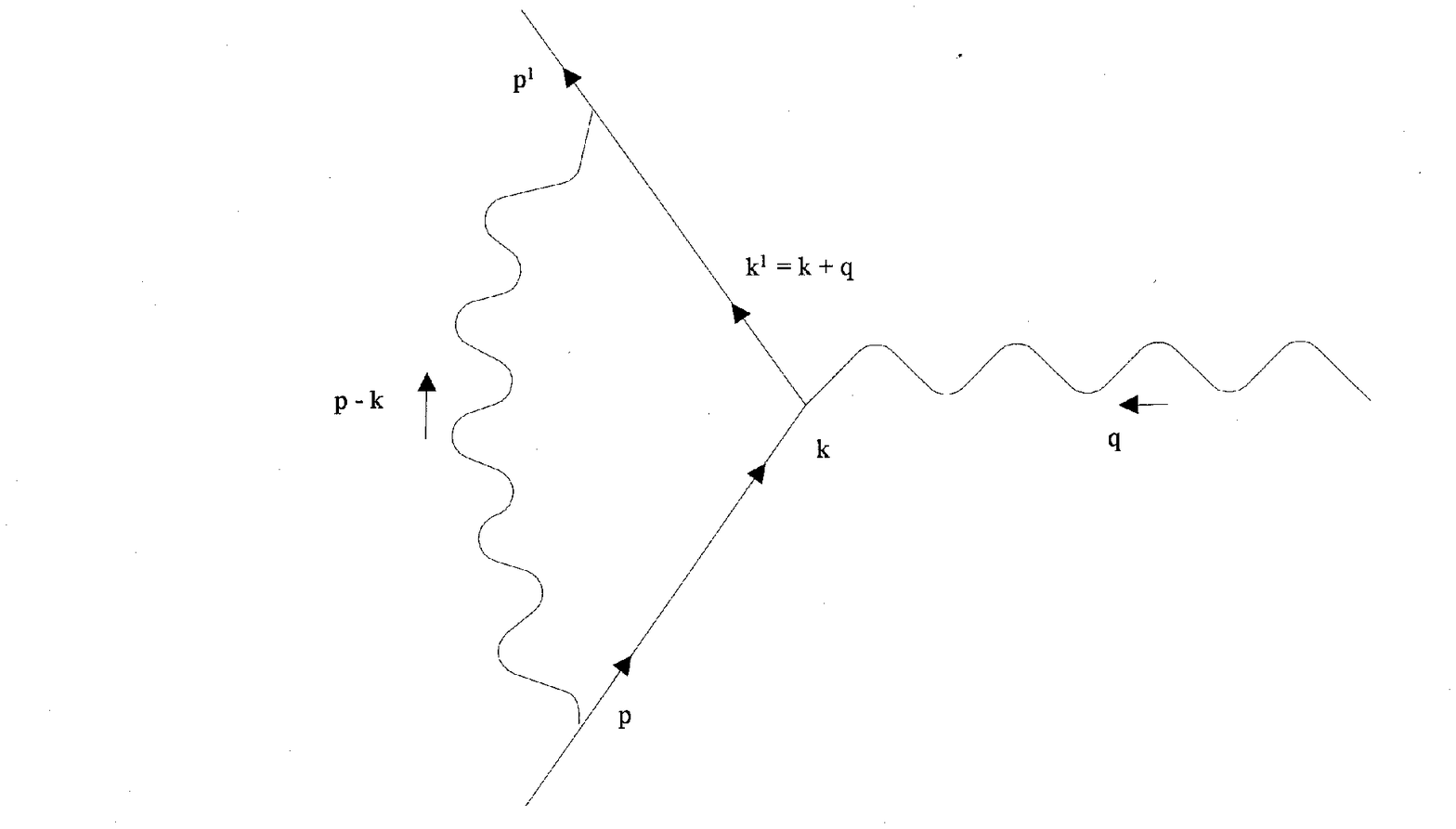}
}\vskip 0.5in
\caption{The electron vertex correction diagram.}
\end{figure}

\end{document}